# Examples for the improvements in AES Depth Profiling of Multilayer Thin Film Systems by Application of Factor Analysis Data Evaluation

U. Scheithauer

Siemens AG, Corporate Research and Development, Otto-Hahn-Ring 6, 81739 München, FRG

Abstract:

Factor Analysis has proved to be a powerful tool for the full exploitation of the chemical information included in the peak shapes and peak positions of spectra measured by AES depth profiling. Due to its ability to extract the number of independent chemical components, their spectra and their depth distributions, its information content exceeds the one of the usual peak-to-peak height evaluation of AES depth profile data. Using modern software with a graphically interactive user interface the analyst is put into a position, where he can work with Factor Analysis on a physically intuitive level despite of all the matrix algebra mathematics which it is based upon.

The progress brought about by Factor Analysis to AES depth profiles of thin films is demonstrated by the analysis of two thin film systems. The first one is a Pt/Ti metallisation used as bottom electrode for ferroelectric thin films, the second one is multilayer system where a Ti silicide formation of burried Ti/Si bilayers has been induced. Both examples show that Factor Analysis evaluation of AES depth profile data is capable to give access to stoichiometry information and to reveal interfacial layer phases, information which is hardly obtained from the conventional peak-to-peak height data evaluation.

## Introduction

In conventional AES depth profile analysis the peak-to-peak heights (ptph) of the first-derivative spectra are plotted as a function of either the sputter time or the ion dose used for sputter erosion. This kind of data evaluation ignores most of the recorded information since only a few data points of the energy intervals measured around the kinetic energy of an AES transition are used. By making full use of all data channels, Factor Analysis (FA) is able to extract the chemical informations included in the Auger peak line shapes and peak positions /1, 2, 3/. In AES depth profiling of thin film systems, FA overcomes the limitations of the conventional first-derivative ptph analysis and yields to not only the depth distributions of elements, but also those of chemical compounds, resolves peak overlaps and often detects interfacial layer phases. This progress is demonstrated by the analysis of two thin film systems.

## Basics of factor analysis:

FA is a multivariant statistical method that uses the all spectral information of the recorded data. It recognises the chemical environment of the atoms from the altered AES peak positions and peak shapes, which themselves depend on the density of electronic states and cross sections. A detailed description of FA is given by Malinowski and Howery /4/. In the following, only the basic concept of the application of FA to AES depth profile measurements shall shortly be recalled.

An AES spectrum measured at a certain sputter depth i can be considered as a vector **D** with r energy grid points. In first-order approximation this spectrum is assumed to be a linear superposition of basic spectra **R**. Utilising this assumption for all sputter depths, the equations can be written in matrix form.

$$[D]_{r \times c} = [R]_{r \times m} [C]_{m \times c} \qquad [1]$$

In each column the data matrix [**D**] contains a spectrum measured at a certain sputter depth. Each column of the matrix [**R**] represents one of the m basic spectra and each row of [**C**] contains the concentrations of one basic spectrum for all c sputter depths. The aim of FA is to determine the basic spectra [**R**] and their concentrations [**C**] from the measured data matrix [**D**].

By means of matrix algebra, the data matrix [**D**] is mathematically decomposed into abstract components [**R**]$^*$ and [**C**]$^*$. The essential information is then separated from noise and modulation of secondary electron background by using only the most significant k principal factors for data reproduction. Only an appropriate choice of k leads to a satisfactory result in the following matrix rotation step.

$$[D]_{r \times c} = [R]^*_{r \times m} [C]^*_{m \times c} \approx [R]^\#_{r \times k} [C]^\#_{k \times c} \qquad [\,2\,]$$

This matrix equation is solved by an infinite number of pairs of matrices [**R**]$^\#$ and [**C**]$^\#$. Generally, the matrix algebra algorithm gives a purely mathematical solution of little useful physical meaning. For example, negative concentrations may occur and the spectra may not look like customary AES spectra. Therefore, the mathematical solution has to be converted into the physically meaningful result by matrix rotations in factor space.

$$[D]_{r \times c} \approx [R]^\#_{r \times k} [T]^{-1} [T] [C]^\#_{k \times c} \qquad [\,3\,]$$

Finding a successful rotation matrix [**T**] is the most crucial and time-consuming step of FA. A novel software has been developed for this purpose allowing the analyst to work with FA on a physically intuitive level despite all complex and abstract mathematics /3/.

The progress brought about by FA to AES depth profile measurements shall now be demonstrated by the analysis of two thin film systems.

## Comparison of an "as deposited" with a thermally treated Pt/Ti metallisation:

Pt/Ti bilayer metallisations are used as bottom electrodes for ferroelectric thin films /5/. In our case, the electrode consists of a 140 nm Pt / 100 nm Ti bilayer deposited by RF sputtering onto a Si wafer which was covered with approximately 500 nm thermally grown oxide. In order to study the effect of the deposition of ferroelectric $Pb(Zr,Ti)O_3$ thin films on this electrode, the deposition process was simulated by an equivalent heat treatment of the electrode in $Ar/O_2$ atmosphere /6/.

The AES depth profiles were acquired using a Perkin-Elmer PHI-600 scanning Auger instrument equipped with a differentially pumped ion gun. For sputter erosion of the sample, 3 keV $Ar^+$ ions with an impact angle of 71° relative to the surface normal were employed. To ensure a good depth resolution the samples were continuously rotated around the surface normal during sputter erosion /7, 8/. AES data were taken from an area of approximately $5 \cdot 10^4$ $\mu m^2$ using a primary electron energy of 10 keV and a primary current of 0.7 µA. The sputtering was interrupted during AES data acquisition.

In Fig. 1 on the l.h.s. the results obtained of the "as deposited" electrode and on the r.h.s. those of the heat treated one are shown. On top of Fig. 1 for each sample the result of a conventional ptph evaluation of the derivative spectra is displayed. The results obtained by factor analysis are depicted in the lower part of Fig. 1.

The data reproduction has been made using 3 and 4 basic factors, respectively. By this choice more about the thin film systems structure has been elaborated than gainable by the ptph data evaluation, a result which justifies the choice. But, of coarse, a FA data reproduction using more basic factors might give additional information, mainly about the interfaces in the case of AES depth profile analysis /3/.

In the middle the derivative eigenspectra in the energy regions of the O KLL, the common one of the Ti LMM (~ 385 eV) and the Ti LMV (~ 418 eV), as well as those of the Si KLL and the Pt MNN transitions are shown. At the bottom of Fig. 1 the relative concentration of these basic factors as function of sputter time are given.

By FA evaluation of the AES depth profile data the basic compounds $SiO_2$, Pt, metallic Ti and two different Ti oxidation states are found. The peak shift of the O KLL signal between O bound to Ti and O bound to Si are becoming obvious. The peak shape differences between Ti in the metallic state and in both oxidation states can also easily be recognised. Considering the ptph of the O KLL signal relative to that of the Ti LMV transition twice the amount of O is found in the basic factor $TiO_y$ than in $TiO_x$. Using these both AES transitions by pure element standard sensitivity factors x and y are estimated to be ~ 1 and ~ 2, respectively. A small amount of TiO with an enrichment at the $Ti/SiO_2$ interface is observed in the Ti layer of the "as deposited" sample. The TiO at the interface is caused by the oxidised surface of the elemental Ti target from which the Ti layer has been sputtered onto the wafer.

Obviously, during the heat treatment the Ti was completely oxidised to form $TiO_2$. This oxidation of the Ti, the correlated uptake of O which thickens the electrode, and the diffusion of Ti into the Pt towards the surface are the most striking differences between the "as deposited" and the heat treated sample. In addition, the electrode/$SiO_2$ interface width increases for the heat treated electrode, thereby indicating an increasing lateral non-uniformity of the electrode.

In summary, in this case FA data evaluation has revealed additionally information about the stoichiometry and the depth distribution of the Ti oxidation states which is not obtainable from ptph data evaluation.

## Silicide formation of buried Si and Ti layers

On a Si substrate a Ti (20 nm) / TiN (100 nm) bilayer, a common diffusion barrier used in the microelectronics technology /9, 10/, was deposited, followed by a Ti and a Si layer, each ~ 100 nm thick. This layer sequence is covered by 300 nm $SiO_2$ on top. To obtain low resistance conductive layers Ti silicide formation of the buried Ti and Si layer has been induced by rapid thermal annealing (RTA). Then, the sample has been characterised by AES depth profiling.

Again, the AES measurements were performed in a Perkin-Elmer PHI-600 scanning Auger instrument equipped with a differentially pumped ion gun. For sputter erosion of the sample, 3 keV $Ar^+$ ions with an impact angle of $55^o$ relative to the surface normal were employed. Sputtering was interrupted during AES data acquisition. The AES data were taken from an area of $\sim 5 \cdot 10^4$ $\mu m^2$ using a primary electron energy of 10 keV and a primary current of 0.1 µA.

In the uppermost portion of Fig. 2 the results of a conventional ptph evaluation of the derivative spectra are depicted. Using the chemical peak shift of Si bound to O towards lower kinetic energy, $SiO_2$ can be discriminated from Si and Si silizide, respectively /11, 12/. The peak minimum of the derivative Si KLL signal at $\sim$ 1609 eV was used to show the depth distribution of $SiO_2$. The minimum of the Si LVV signal at $\sim$ 92 eV corresponds to both, Si and Si silizide. Since the N KLL signal, the only Auger transitions of N, coincides with the Ti LMM transitions at $\sim$ 385 eV, a ptph evaluation of the data can only give the depth distribution of the sum signal.

With FA four basic factors were chosen to reproduce the measured data sufficiently precisely. These factors are identified as Si, $SiO_2$, Ti-Si and Ti-N. In the middle of Fig. 2 the rotated derivative eigenspectra are displayed. The energy region of the O KLL transition, the common one of the N KLL, Ti LMM and Ti LMV transitions as well as those of the Si LVV and Si KLL Auger signals are shown. At the bottom the corresponding relative concentrations of these spectra are plotted as a function of sputter time. As can be seen, the expected RTA induced silizide formation of the buried Si and Ti layer has taken place, but some unreacted Si remains on top of the silizide. The Ti layer of the diffusion barrier has reacted with the Si substrate to form a silizide, too. A small amount of Ti silizide is found within the TiN. The Auger spectra belonging to the four basic factors show the typical peak shifts of the Si signals as discussed above. Additionally the analyst is supplied with the exact peak shapes of the Auger signals belonging to the different material compounds recorded with the actual analyser transfer function, for instance with those of Ti silizide and Ti nitride.

In comparison to the ptph depth profile, the relative concentration depth profiles of the material compounds are obtained from the FA evaluation, and thus give a much better insight into the Ti silizide formation process of the annealed thin film system.

## Conclusions:

In contrast to conventional ptph analysis the multivariant statistical method FA makes full use of all the recorded data. It gives access to the chemical information included in the Auger peak line shapes and peak positions. As demonstrated by the two analysed thin film systems overlapping peaks are resolved, stoichiometric informations are obtained, and interface reaction zones are detected that are usually easily overlooked. Generally, the progress brought about by FA to AES depth profile data evaluation consists in obtaining depth distributions of chemical compounds with refined information about stoichiometry and interfacial layers.

## Acknowledgement

The samples where kindly provided by R. Bruchhaus and G. Tempel from Siemens Corporate Research and Development Laboratories. The software code is commercially available through Synotec GmbH, In der Zeil 39, 91058 Erlangen, FRG.

## Literature:

/1/   J.S. Solomon, Surf. Interface Anal., 10, 75-86 (1987)

/2/   S.W. Gaarenstroom, Appl. Surf. Sci., 26, 561-574 (1986)

/3/   U. Scheithauer, W. Hösler and G. Riedl, Surf. Interface Anal., 20, 519-523 (1993)

/4/   E. R. Malinowski and D. G. Howery, Factor Analysis in Chemistry, John Wiley & Sons, New York, Chichester, Brisbane, Toronto (1980)

/5/   R. Bruchhaus, D. Pitzer, O. Eibl, U. Scheithauer and W. Hösler, Mater. Res. Soc. Proc. 243, 123-128 (1992)

/6/   U. Scheithauer, W. Hösler and R. Bruchhaus, Fresenius J. Anal. Chem., 346, 305-307 (1993)

/7/   K. Vogelbruch and E. Vietzke, Symp. on Fusion Technology, 829-833 (1982), Pergamon Press, Oxford, New York

/8/   A. Zalar, Thin Solid Films 124, 223-230 (1985)

/9/   J.R. Shappirio, Solid State Technology, Oct. 1985, 161-166 (1985)

/10/  D. Widmann et.al., Technologie hochintegrierter Schaltungen, Springer, Berlin, Heidelberg, New York, 1988

/11/  L.E. Davis (1978) Handbook of Auger Spectroscopy, Perkin Elmer

/12/  C. Quenisset, R. Naslain and P. Demoncy, Surf. Interface Anal., 13, 123-129 (1988)

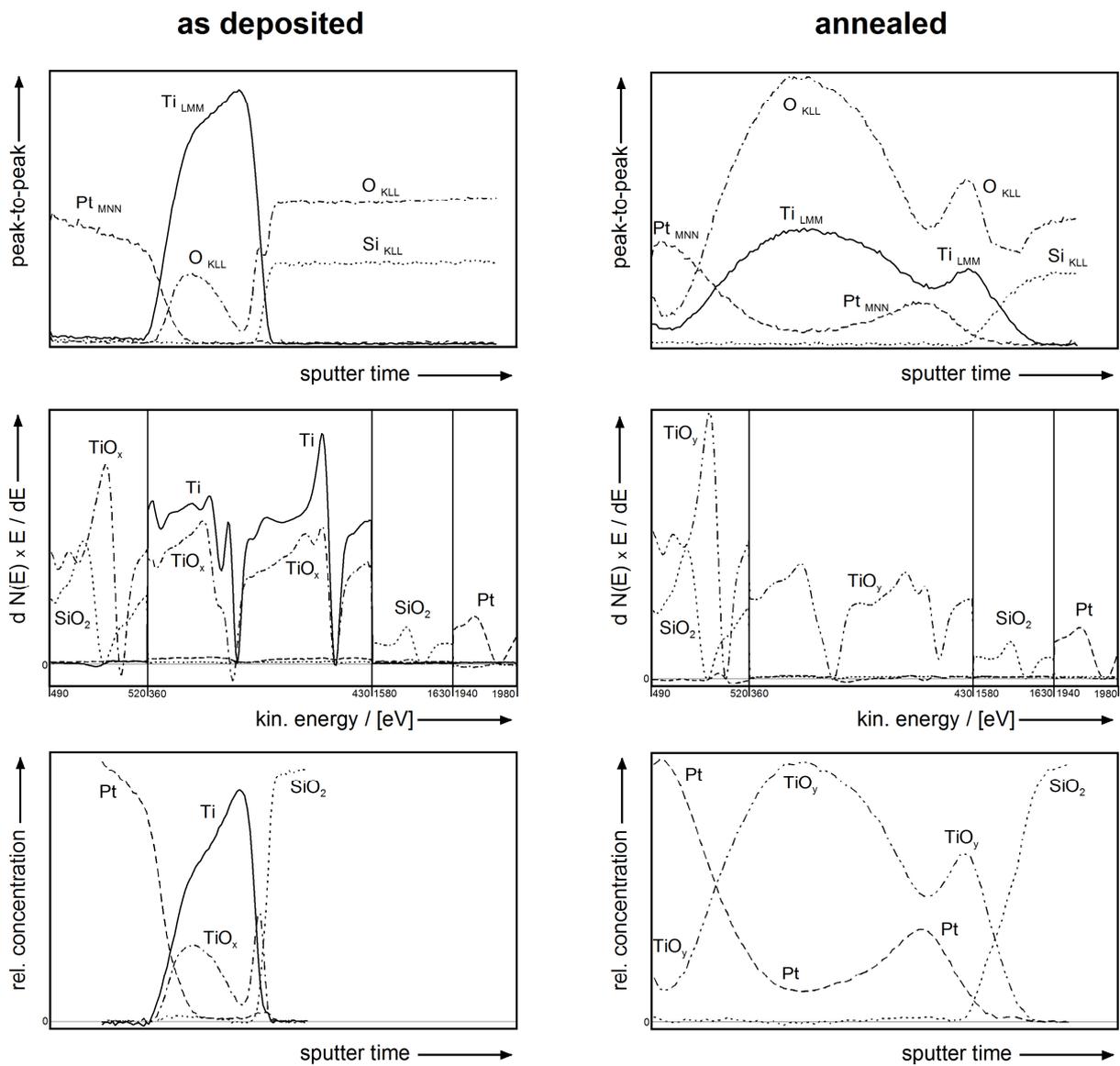

Fig. 1: Pt / Ti electrode for ferroelectric thin films, comparison of a "as deposited" with an annealed electrode

(20 min ramped up to T = 450°C and 120 min annealed in 1,4 Pa Ar/$O_2$ atmosphere)

top: AES peak-to-peak heights data evaluation of first-derivative spectra,

middle: basic spectra found by Factor Analysis,

bottom: relative concentrations of basic spectra found by Factor Analysis

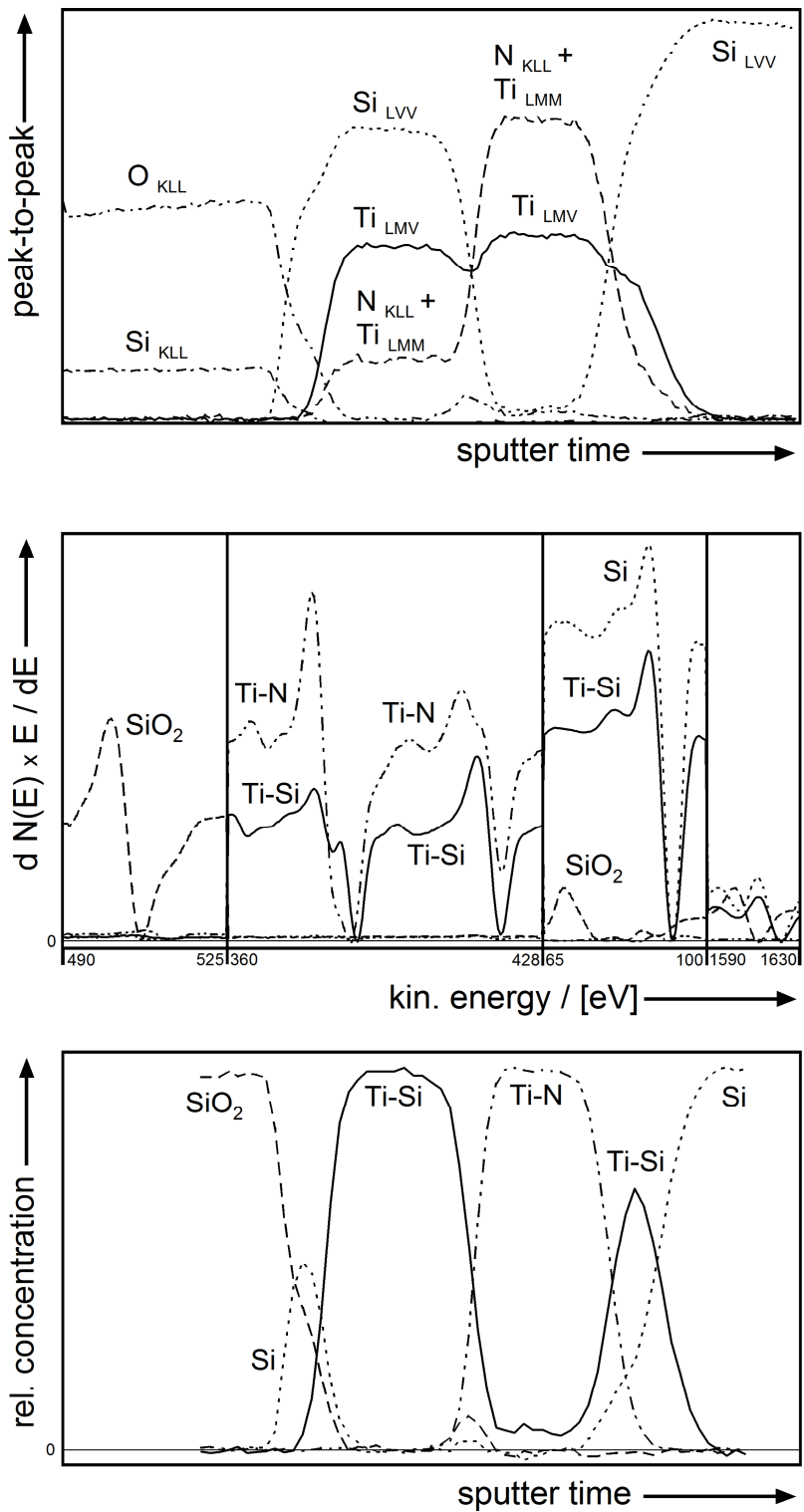

Fig. 2: silicide formation of burried Si and Ti layers by RTA

    top: AES peak-to-peak heights data evaluation of first-derivative spectra,

    middle: basic spectra found by Factor Analysis,

    bottom: relative concentrations of basic spectra found by Factor Analysis